\documentclass[aps,prd,preprintnumbers,groupedaddress,nofootinbib,amssymb,notitlepage,eqsecnum]{revtex4-2}
\usepackage{here}
\usepackage{comment}
\usepackage[dvipdfmx]{graphicx}
\usepackage{amsmath,amsthm,amssymb}
\usepackage{bm}
\usepackage{color}
\usepackage{comment}
\usepackage{simplewick}

\usepackage{amsfonts}
\usepackage{dcolumn}
\usepackage{hyperref}
\allowdisplaybreaks[1]
\usepackage{stackengine}

\newcommand{\be}{\begin{equation}}  
\newcommand{\ee}{\end{equation}}
\newcommand{\ba}{\begin{eqnarray}}
\newcommand{\ea}{\end{eqnarray}}

\newcommand{\bem}{\begin{bmatrix}}
\newcommand{\eem}{\end{bmatrix}}

\newcommand{\ri}{{\rm i}}

\allowdisplaybreaks

\begin{document}

\preprint{WUCG-24-03}

\title{Roles of boundary and equation-of-motion 
terms in cosmological correlation functions}

\author{Ryodai Kawaguchi$^1$\footnote{{\tt ryodai0602@fuji.waseda.jp}}}
\author{Shinji Tsujikawa$^1$\footnote{{\tt tsujikawa@waseda.jp}}}
\author{Yusuke Yamada$^2$\footnote{{\tt y-yamada@aoni.waseda.jp}}}

\affiliation{${}^1$
Department of Physics, Waseda University, 3-4-1 Okubo, Shinjuku, Tokyo 169-8555, Japan}

\affiliation
{${}^2$
Waseda Institute for Advanced Study, Waseda University, 1-21-1 Nishi Waseda, Shinjuku, Tokyo 169-0051, Japan}

\begin{abstract}

We revisit the properties of total time-derivative terms 
as well as terms proportional to 
the free equation of motion (EOM) in a 
Schwinger-Keldysh formalism. They are relevant to the 
correct calculation of correlation functions of curvature perturbations in the context of 
inflationary Universe. 
We show that these two contributions to the action 
play different roles in the operator or the path-integral formalism, but they give the same correlation functions as each other. As a concrete example, we confirm that the 
Maldacena's consistency relations for the three-point 
correlation function in the slow-roll inflationary scenario 
driven by a minimally coupled canonical scalar field 
hold in both the operator and path-integral formalisms.
We also give some comments on loop calculations.

\end{abstract}

\date{\today}

%\pacs{04.50.Kd, 95.36.+x, 98.80.-k}

\maketitle

%%%%%%%%%%%%%%%%%%%%%%%%%%%%%%%%%%%%%%%%%%
\section{Introduction}
\label{introduction}
%%%%%%%%%%%%%%%%%%%%%%%%%%%%%%%%%%%%%%%%%%

The cosmological perturbation theory has been 
developed as a tool to study the prediction e.g., 
of inflation models 
in the early Universe \cite{Bardeen:1980kt,Kodama:1984ziu,Mukhanov:1990me}.
The spectra of primordial perturbations generated during 
inflation can be tested via cosmological observations such as cosmic microwave background temperature 
anisotropies \cite{Planck:2018vyg,Planck:2018jri}. 
In inflationary Universe, the quantum origin of those perturbations should be addressed by 
quantum field theory (QFT) on a quasi 
de-Sitter background (see e.g., Ref.~\cite{Weinberg:2005vy}). 
One of the differences from the S-matrix theory for the scattering of particles in Minkowski spacetime is the use of a Schwinger-Keldysh (or the in-in or the expectation value) formalism, as we evaluate the expectation values of operators rather than transition amplitudes. 

In the Schwinger-Keldysh formalism, there is a notion of 
the final spacelike surface at which we often insert operators 
that we evaluate. The presence of such a final surface does not allow us to freely neglect total time-derivative terms in the Lagrangian since they appear in general as boundary operators. Furthermore, in inflationary models, it is often convenient 
to use integration-by-part techniques to reduce the action such that the interaction vertices are proportional to powers of small slow-roll parameters, which makes clear the validity of the perturbative approach. 
Such a procedure leaves the aforementioned total 
time-derivative terms as well as terms that 
are proportional to the free equation of motion (EOM) 
of fields in the Lagrangian, which we call the EOM terms. In the first result of QFT calculations of the three-point correlation function of cosmological perturbations performed by Maldacena~\cite{Maldacena:2002vr}, he used a field redefinition technique to remove both the boundary and the EOM terms simultaneously.

In this paper, we will revisit the question of how the boundary and EOM terms play their roles in the practical 
calculations of the cosmological correlation functions. It has been known 
that, even without performing the field redefinition, the boundary terms give the same result 
as the field redefinition while it is often claimed 
that the EOM terms play no
roles~\cite{Seery:2006tq,Chen:2006nt,Arroja:2011yj,Burrage:2011hd}.
However, as we will show below, such a claim is correct only within the operator formalism but not within 
the path-integral formalism.
In the path-integral formalism, the boundary terms do not contribute to the correlation function 
by setting the final slice time $\tau_f$ later than $\tau_*$ 
at which the expectation value of operators is evaluated, but the EOM terms instead become nonnegligible, 
with which we find the same result as the operator formalism.

One of the reasons why we address such a rather technical issue is the recent discussion regarding the presence or absence of large one-loop corrections to the scalar perturbation 
in inflationary models with a short ultra-slow-roll phase 
between two slow-roll regimes~\cite{Kristiano:2022maq,Riotto:2023hoz,Riotto:2023gpm,Kristiano:2023scm,Choudhury:2023vuj,Choudhury:2023jlt,Choudhury:2023rks,Firouzjahi:2023aum,Motohashi:2023syh,Firouzjahi:2023ahg,Franciolini:2023lgy,Tasinato:2023ukp,Cheng:2023ikq,Fumagalli:2023hpa,Maity:2023qzw,Tada:2023rgp,Iacconi:2023ggt,Firouzjahi:2024psd,Inomata:2024lud}. The core of the argument of Ref.~\cite{Kristiano:2022maq} is that there is a possibly large coupling $\eta'\zeta'\zeta^2/2$ 
in the cubic-order action, 
where $\eta$ is one of the slow-roll parameters and
a prime denotes the derivative with respect to 
conformal time. This can generate a large one-loop contribution to the two-point correlation function around the time of phase transitions at which $|\eta'|$ becomes very large. 

On the other hand, in Refs.~\cite{Fumagalli:2023hpa,Tada:2023rgp}, it was argued that including contributions from some total 
time-derivative terms cancels seemingly large contributions from the above coupling. Furthermore, it was shown that 
Maldacena's consistency relation can be reproduced only when the total time-derivative terms are taken into account \cite{Motohashi:2023syh,Tada:2023rgp}. 
The consistency relation can also be proven within the field redefinition method~\cite{Kristiano:2023scm} in the transient 
ultra-slow-roll model if one properly takes into account the additional contributions associated with field redefinition. 
In light of such observations, it would be worthwhile to 
have a better understanding of the roles of boundary and EOM terms for the computation of higher-order corrections. 
Note also that the coupling $\eta'\zeta'\zeta^2/2$ in the bulk action is a consequence 
of the repeated applications of integration-by-part techniques, which yields multiple total 
time-derivative terms. Hence, one expects that the bulk and the ``boundary terms'' are not independent of each other.

The rest of the paper is organized as follows. 
In Sec.~\ref{boundsec}, we discuss how the boundary terms and the EOM terms contribute to the correlation 
functions within the operator and the path-integral formalisms, respectively. 
In Sec.~\ref{examples}, we then illustrate the roles 
of those terms by computing three-point correlation functions 
and show that they may play important roles in reproducing consistency relations for various correlation functions. 
We also give a few comments on higher-order corrections 
in Sec.~\ref{Higherorder}. 
Sec.~\ref{summary} is devoted to conclusions. 
In appendix~\ref{ambiguityId}, we note an identity with which the bulk, boundary, and EOM terms can be changed. Appendix~\ref{noteonBT} is almost independent of the other part, and we discuss some issues of total 
time-derivative terms in the Hamiltonian, in general.

%%%%%%%%%%%%%%%%%%%%%%%%%%%%%%%%%%%%%%%%%
\section{Roles of boundary and EOM terms in Schwinger-Keldysh formalism}
\label{boundsec}
%%%%%%%%%%%%%%%%%%%%%%%%%%%%%%%%%%%%%%%%%

In this paper, we are interested in the correlation 
functions of scalar perturbations on a spatially-flat 
Friedmann-Lema\^{i}tre-Robertson-Walker (FLRW) background 
given by the line element 
$ds^2=-dt^2+a^2(t) \delta_{ij}dx^i dx^j$, where 
$a(t)$ is the scale factor that depends on 
the cosmic time $t$.
In particular, we consider a canonical scalar field $\phi$ minimally coupled to gravity, which is 
described by the action 
\be
S=\int d^4 x \sqrt{-g} \left[ \frac{M_{\rm Pl}^2}{2}R 
-\frac{1}{2}\partial_{\mu} \phi \partial^{\mu} \phi
-V(\phi) \right]\,,
\label{action}
\ee
where $M_{\rm Pl}$ is the reduced Planck mass, $R$ is 
the Ricci scalar, and $V(\phi)$ is a scalar 
potential. The scalar field has a perturbation 
$\delta \phi (t, x^i)$ on the time-dependent 
background value $\phi(t)$. 
For the perturbed line element, we choose the 
Arnowitt-Deser-Misner metric expressed in the form 
\be
ds^2=-N^2 dt^2+a^2(t) e^{2\zeta (t,x^i)}
\delta_{ij} \left( dx^i+N^i dt 
\right) \left( dx^j+N^j dt \right)\,,
\ee
where $N$ is the lapse, $N^i$ is the shift, and 
$\zeta$ is the curvature perturbation. 
We choose the unitary gauge $\delta \phi=0$ and 
integrate out $N$ and $N^i$ from the action 
by using their constraint equations of motion.

Expanding the action up to the second order in perturbations 
and varying it with respect to $\zeta$, we obtain 
the following linear perturbation equation
\be
(a^2\epsilon \zeta')'
-a^2\epsilon\,\partial^2 \zeta=0\,,
\label{lineareq}
\ee
where a prime represents the derivative with 
respect to the conformal time $\tau=\int a^{-1}dt$, 
and 
\be
\epsilon=-\frac{\dot{H}}{H^2}\,,\qquad 
H=\frac{\dot{a}}{a}\,,
\ee
with a dot being the derivative with respect to $t$. 
If we consider the scalar field $\phi$ as an inflaton, the curvature perturbation 
exits the Hubble horizon during inflation. 
Far outside the Hubble radius, 
the spatial gradient term $\partial^2 \zeta$ in 
Eq.~(\ref{lineareq}) can be neglected relative to 
the first term. In this regime, we obtain the
following solution for the linear 
curvature perturbation
\be
\zeta=c_1+c_2 \int \frac{d \tau}{a^2 \epsilon}\,,
\label{zetaso}
\ee
where $c_1$ and $c_2$ are integration constants. 
In standard slow-roll inflation where $\epsilon$ 
slowly varies in time, 
the second term on the 
right-hand side (RHS) of Eq.~(\ref{zetaso}) 
corresponds to a decaying mode. In this case, 
$\zeta$ approaches a constant $c_1$ 
after the Hubble radius 
crossing. We note that, in the context of 
ultra-slow-roll inflation where there is a temporal 
epoch in which $\epsilon$ rapidly decreases toward 0, 
the second term in Eq.~(\ref{zetaso}) can lead to 
the enhancement of $\zeta$. 

In this paper, we are primarily interested in 
the nonlinear curvature perturbation and 
its three-point correlation function in the 
context of slow-roll inflation.
For this purpose, the cubic-order action 
of $\zeta$, whose explicit form will be presented 
in Sec.~\ref{examples}, plays a crucial role.
Such a cubic action contains boundary terms as well as terms 
proportional to the left-hand side 
of the linear EOM (\ref{lineareq}). 
Note that the latter is generally 
nonvanishing for nonlinear perturbations. 
From the viewpoint of QFT, we need to properly 
deal with these two contributions to the action 
for calculating the three-point correlation function 
as well as loop corrections.
In this section, we will explain two general 
frameworks for computing the correlation functions 
in the presence of the boundary and EOM terms. 

%%%%%%%%%%%%%%%%%%%%%%%%%%%%%%%%%%%%%
\subsection{The operator formalism}
%%%%%%%%%%%%%%%%%%%%%%%%%%%%%%%%%%%%%%%%%

We begin with the operator formalism for dealing with 
the boundary terms in the action.
Note that the role of such contributions 
has been discussed in the 
literature~\cite{Seery:2006tq,Arroja:2011yj,Burrage:2011hd}. 
For concreteness, we consider the following interaction Hamiltonian of curvature perturbations
\begin{align}
\hat{H}_{\rm int}=\hat{H}_{\rm bulk}+\int d^3\bm x \left[({\cal{D}}_2 \hat{\zeta})\hat f(\tau)+\frac{d}{d\tau}\hat{g}(\tau)\right]\,,
\label{interactionH}
\end{align}
where $\hat{H}_{\rm bulk}$ is the Hamiltonian 
in the bulk, 
$\hat{f}(\tau)$ and $\hat{g}(\tau)$ are arbitrary 
time-dependent functions of the operator $\hat{\zeta}$ 
possibly including derivative terms, and we have defined 
\be
{\cal{D}}_2 \hat{\zeta} \equiv 2M_{\rm Pl}^2 \left[
(a^2\epsilon \hat{\zeta}')'
-a^2\epsilon\,\partial^2 \hat{\zeta} \right]\,.
\ee
The first contribution to the integrand in 
Eq.~(\ref{interactionH}) is the EOM term and 
the second one corresponds to the boundary term.
Note that, in the spacetime that has spatial translation invariance such as the FLRW spacetime, 
the spatial total derivative terms are simply neglected without any subtlety thanks to the three-momentum conservation.

In the operator formalism, the interaction picture operator $\hat\zeta^I(\tau,\bm x)$ consists of the creation and annihilation operators multiplied by either positive or negative frequency solutions to the free equation of motion ${\cal{D}}_2\zeta=0$. 
Therefore, we can drop the contribution of EOM terms from the interaction Hamiltonian operator independently of the function $\hat{f}(\tau)$. 
Thus, we will focus only on the total time derivative term in the Hamiltonian. 
As we will discuss later, it is worth emphasizing that this is not the case 
in path-integral formalism where the derivatives appearing interactions act on the propagators.

Secondly, we should note that the total time derivative cannot be simply integrated 
within the time-dependent perturbation theory. 
The time-evolution operator is given by
\begin{align}
\hat{U}(\tau,\tau_0)=T\exp\left[-\ri \int_{\tau_0}^\tau d\tilde{\tau} \hat{H}_{\rm int}(\tilde{\tau})\right]\,,
\end{align}
and one may think that the total time-derivative operator can be simply integrated as
\begin{align}
\hat{U}(\tau,\tau_0)\overset{?}{=}e^{-\ri \int d^3\bm x \hat{g}(\tau)}T\exp\left[-\ri\int d^3\bm x \int_{\tau_0}^\tau d\tilde{\tau} \hat{H}_{\rm bulk}(\tilde{\tau})
\right]e^{\ri \int d^3\bm x \hat{g}(\tau_0)}\,.\label{Uexpect}
\end{align}
However, the RHS is simply wrong unless $\hat{g}(\tau)$ commutes with bulk terms $\hat{H}_{\rm bulk}(\tau)$. Thus, the total time-derivative term cannot really be treated as the boundary term at the final surface\footnote{If the time integration is naively carried out, the total time derivative terms can be regarded as the operators at the ``initial'' and the ``final'' Cauchy slices. Assuming that interactions are turned off at the initial time, we would be left with the operators on the final surface. It is often assumed by the Bunch-Davies vacuum condition (or $\ri \epsilon$ prescription or Euclidean vacuum condition).}. 
The correct expression when $\hat{H}_{\rm bulk}(\tau)$ is included is
\begin{align}
\hat{U}(\tau,\tau_0)=&\hat{\bm 1}-\ri\int_{\tau_0}^\tau d\tau_1\left[ \hat{H}_{\rm bulk}(\tau_1)+\frac{d}{d\tau_1}\hat{g}(\tau_1) \right]
\nonumber\\
&+(-\ri)^2\int^{\tau}_{\tau_0}d\tau_1\int^{\tau_1}_{\tau_0}d\tau_2\left[ \hat{H}_{\rm bulk}(\tau_1)+\frac{d}{d\tau_1}\hat{g}(\tau_1)\right] 
\left[\hat{H}_{\rm bulk}(\tau_2)+\frac{d}{d\tau_2}\hat{g}(\tau_2)\right]+\cdots\,,
\end{align}
where we have formally introduced the bulk interaction Hamiltonian and the ellipses denote higher-order terms. Now, it is clear that the total time-derivative terms cannot be simply integrated starting from the second order in the interaction Hamiltonian. Nevertheless, at first order, one can perform the time integration and obtain
\begin{align}
\hat{U}(\tau,\tau_0)|_{\rm 1st}=\hat{\bm 1}-\ri \hat{g}(\tau)-\ri\int_{\tau_0}^\tau d\tau_1\hat{H}_{\rm bulk}(\tau_1),
\end{align}
where we have assumed $\hat{g}(\tau_0)=0$, and the total time derivative term 
contributes to the correlation function of operators 
$\langle\hat{O}_1\hat{O}_2\rangle$ as
\begin{align}
\langle\hat{O}_1(\tau_*)\hat{O}_2(\tau_*)\rangle\sim \langle\hat{O}^I_1(\tau_*)\hat{O}^I_2(\tau_*)\rangle+\ri \int_{\tau_0}^{\tau_*} d\tau\langle [\hat{H}^I_{\rm bulk}(\tau),\hat{O}^I_1(\tau_*)\hat{O}^I_2(\tau_*)]\rangle+\ri\langle[\hat{g}(\zeta^I(\tau_*)),\hat{O}^I_1(\tau_*)\hat{O}^I_2(\tau_*)]\rangle,\label{boundary1st}
\end{align}
at first order in the interaction 
Hamiltonian (the operator with ``$I$'' denotes the interaction picture operator). Indeed, the proper modification of \eqref{Uexpect} can be found in Refs.~\cite{Weinberg:2005vy,Braglia2024}, which is relatively concise. We will not discuss it here, but see Ref.~\cite{Braglia2024} for careful and detailed derivations of the corresponding formula. 

Let us illustrate how the boundary Hamiltonian plays its role in the computation of correlation functions. 
We consider a simple boundary Hamiltonian $\hat{g}(\tau)=\int d^3\bm x A(\tau)\hat{\zeta}(\tau,\bm x)\hat{\pi}_\zeta(\tau,\bm x)\hat{\zeta}(\tau,\bm x)$, where $A(\tau)$ is a real function and $\hat{\pi}_\zeta$ denotes the canonical conjugate momentum of $\hat{\zeta}$. For a three-point correlation function $\langle\hat{\zeta}(\tau_*,\bm x_1)\hat{\zeta}(\tau_*,\bm x_2)\hat{\zeta}(\tau_*,\bm x_3)\rangle$ we find, from the last term of Eq.~\eqref{boundary1st}, 
\ba
\langle\hat{\zeta}(\tau_*,\bm x_1)\hat{\zeta}(\tau_*,\bm x_2)\hat{\zeta}(\tau_*,\bm x_3)\rangle 
&\supset& 
\ri A(\tau_*)\int d^3\bm x\langle[\hat{\zeta}^I(\tau_*,\bm x)\hat{\pi}_\zeta^I(\tau_*,\bm x)\hat{\zeta}^I(\tau_*,\bm x),\hat{\zeta}^I(\tau_*,\bm x_1)\hat{\zeta}^I(\tau_*,\bm x_2)\hat{\zeta}^I(\tau_*,\bm x_3)] \rangle \nonumber\\    &=&\ri A(\tau_*)\int d^3\bm x\langle\hat{\zeta}^I(\tau_*,\bm x) [\hat{\pi}_\zeta^I(\tau_*,\bm x),\hat{\zeta}^I(\tau_*,\bm x_1)\hat{\zeta}^I(\tau_*,\bm x_2)\hat{\zeta}^I(\tau_*,\bm x_3)] \hat{\zeta}^I(\tau_*,\bm x)\rangle\nonumber\\
&=& A(\tau_*)\langle\hat{\zeta}^I(\tau_*,\bm x_2)\hat{\zeta}^I(\tau_*,\bm x_3)\left(\hat{\zeta}^I(\tau_*,\bm x_1)\right)^2+({\rm cyclic})\rangle\nonumber\\
&\to&A(\tau_*)\langle \hat{\zeta}^I(\tau_*,\bm x_1)\hat{\zeta}^I(\tau_*,\bm x_2)\rangle\langle\hat{\zeta}^I(\tau_*,\bm x_1)\hat{\zeta}^I(\tau_*,\bm x_3)\rangle+(\rm cyclic)\,,
\ea
where ``$(\rm cyclic)$'' denotes the cyclic permutation with respect to $(1,2,3)$ and the last line shows only the connected part of the correlation function. 
Here, we have used the canonical commutation relation in the third line, from which we find the following: The equal time commutation relation yields the spatial $\delta$-function, and an operator ``contracted'' by the commutation relation is effectively ``redefined''. This behavior is nothing but the field redefinition prescription proposed by Maldacena~\cite{Maldacena:2002vr}. Indeed, we could reach the same expression by performing the field redefinition $\zeta\to \zeta+A\zeta^2$. This point has been pointed out in Refs.~\cite{Seery:2006tq,Arroja:2011yj,Burrage:2011hd}. As we will see in the next subsection, the role of the ``boundary terms'' is played instead by EOM terms 
within the path-integral formalism.

%%%%%%%%%%%%%%%%%%%%%%%%%%%%%%%%%%%%%%%%%%
\subsection{The path-integral formalism}
\label{pathintegral}
%%%%%%%%%%%%%%%%%%%%%%%%%%%%%%%%%%%%%%%%%%

In this subsection, we clarify how the boundary 
and the EOM terms contribute to the evaluation of the cosmological correlation functions in the 
path-integral formalism, which looks different from that in the operator formalism.
In the operator approach, we take the expectation value of interaction picture operators that possibly contain derivatives, 
whereas, in the path-integral approach, correlation functions consist of free propagators, and the derivatives contained in the interaction vertices would act from outside of the propagator. In other words, we do not evaluate operators with derivatives, but differentiate propagators when vertices contain differential operators. 
Then, we cannot ignore the terms proportional to the EOM as the (anti) time-ordered products 
do not vanish on-shell, while non-time-ordered products do. 
Therefore, in the Schwinger-Keldysh path integral, 
the EOM term should not be neglected.

Let us briefly review the Schwinger-Keldysh path-integral formalism. 
See e.g., Refs.~\cite{Weinberg:2005vy,Chen:2017ryl} for detailed discussions. As we evaluate expectation values rather than S-matrix elements, there are ``bra'' and ``ket'' 
fields\footnote{They are necessary for anti-time ordered and time ordered evolution associated with bra and ket vectors.}  $\zeta^-$ and $\zeta^+$, 
respectively, and the degrees of freedom are doubled in comparison to the standard Feynman path integral. 
The free generating functional of the Schwinger-Keldysh 
path integral is given by
\begin{align}
Z_0[J^+,J^-]=\mathcal{N}_0\int {\mathcal D}\zeta^+\mathcal{D}\zeta^-\delta(\zeta^+(\tau_f,\bm x)-\zeta^-(\tau_f,\bm x))\exp\left[-\ri S_2[\zeta^-]+\ri S_2[\zeta^+]+\ri\int d\tau d^3{\bm x}(J^+\zeta^+-J^-\zeta^-)\right],
\end{align}
where $\tau_f$ denotes the final slice time, 
$J^\pm$ are source functions, and $\mathcal{N}_0$ 
is a normalization factor, and 
$S_2[\zeta]$ denotes the quadratic action of $\zeta$ given by
\be
S_2[\zeta]=M_{\rm Pl}^2 \int d\tau d^3 {\bm x}\, a^2 \epsilon 
\left[ \zeta'^2 -(\partial \zeta)^2 \right]\,.
\ee
The $\delta$-function enforces the field configurations of $\zeta^\pm$ to coincide at $\tau=\tau_f$. 
The partition function of the interacting theory can be perturbatively calculated by
\begin{align}
Z[J^+,J^-]=\mathcal{N}\exp\left[\ri S_{\rm int}\left[\frac{\delta}{\ri \delta J^+}\right]-\ri  S_{\rm int}\left[\frac{\delta}{-\ri \delta J^-}\right]\right]Z_0[J_+,J_-]\,,
\end{align}
where $S_{\rm int}$ denotes the interacting part of the action possibly including the spacetime 
derivative of $\zeta$. 
The connected correlation function can be computed 
via the generating functional
$W[J^+,J^-]=-\ri \log \left(Z[J^+,J^-]\right)$, with which the three-point correlation function, 
for example, is given by
\begin{align}
\langle\zeta^+(\tau_*,\bm x_1)\zeta^+(\tau_*,\bm x_2)\zeta^+(\tau_*, \bm x_3)\rangle=\frac{\delta^3}{\ri^3\delta J^+(\tau_*,\bm x_1)\delta J^+(\tau_*,\bm x_2)\delta J^+(\tau_*,\bm x_3)}W[J^+,J^-]\biggr|_{J=0}\,.
\end{align}
The evaluation of the correlation functions can be done by contracting the interaction vertices and the operators put at $\tau=\tau_*$, and due to the presence of the $\pm$-fields, we have four different propagators:
\begin{align}
G^{++}(\tau_1,\bm x_1;\tau_2,\bm x_2)=&\frac{\delta^2}{\ri^2 \delta J^+(\tau_1,\bm x_1)\delta J^+(\tau_2,\bm x_2)}Z_0[J^+,J^-]\biggr|_{J=0}=\langle T(\hat{\zeta}(\tau_1,\bm x_1)\hat{\zeta}(\tau_2,\bm x_2))\rangle,\\
G^{+-}(\tau_1,\bm x_1;\tau_2,\bm x_2)=&\frac{\delta^2}{ \delta J^+(\tau_1,\bm x_1)\delta J^-(\tau_2,\bm x_2)}Z_0[J^+,J^-]\biggr|_{J=0}=\langle \hat{\zeta}(\tau_2,\bm x_2)\hat{\zeta}(\tau_1,\bm x_1)\rangle,\\
G^{-+}(\tau_1,\bm x_1;\tau_2,\bm x_2)=&\frac{\delta^2}{ \delta J^-(\tau_1,\bm x_1)\delta J^+(\tau_2,\bm x_2)}Z_0[J^+,J^-]\biggr|_{J=0}=\langle \hat{\zeta}(\tau_1,\bm x_1)\hat{\zeta}(\tau_2,\bm x_2)\rangle,\\
G^{--}(\tau_1,\bm x_1;\tau_2,\bm x_2)=&\frac{\delta^2}{(-\ri)^2 \delta  J^-(\tau_1,\bm x_1)\delta J^-(\tau_2,\bm x_2)}Z_0[J^+,J^-]\biggr|_{J=0}=\langle \bar{T}(\hat{\zeta}(\tau_1,\bm x_1)\hat{\zeta}(\tau_2,\bm x_2))\rangle\,,
\label{propagators}
\end{align}
where $T(\bar{T})$ is (anti-)time ordering symbol. 

One question arises here.
How should the time $\tau_f$ be chosen for the final slice, which is introduced to calculate 
the expectation value in the path-integral formalism?
As long as $\tau_f\ge\tau_*$ is satisfied, the choice is arbitrary and should not affect the final result.
In general, $\tau_f=\tau_*$ has been the preferred choice. This is because, since we are mostly interested in correlation functions of $\zeta$ without derivatives at $\tau_*$, we can freely take them to be either $+$ or $-$ fields thanks to the $\delta$-function in 
the partition function.
However, the $\delta$-function $\delta(\zeta^+(\tau_f)-\zeta^-(\tau_f))$ that enforces the bra- and ket-fields $\zeta^\mp$ to be equal is not necessary by the following reason: In cosmological applications, we are mostly interested in the equal time correlation functions of the same operator $\langle\zeta(\tau_*,\bm x_1)\cdots\zeta(\tau_*,\bm x_n)\rangle$. In such cases, they are products of commuting operators and their ordering does not really change anything. 
Thus, they can be identified as either $+$ or $-$ fields\footnote{Even if one wants to evaluate the expectation value of a product of operators that are not commutative, the ordering ambiguity does not contribute to the connected diagram.}.
Therefore, in the following, we choose $\tau_f>\tau_*$, which makes unambiguous the contributions from 
the EOM terms and the boundary terms.
As we will see later, the EOM terms lead to a $\delta$-function, $\delta(\tau-\tau_*)$, 
so it is more healthy to take the integral range of $\tau$ as fully encompassing $\tau_*$.

Now, let us consider the following action
\begin{align}
S_{\rm int}=-\int d\tau d^3\bm x \left[(\mathcal{D}_2\zeta) f(\zeta)+\frac{d}{d\tau}g( \zeta)\right]\,,
\end{align}
which describes the interaction.
One of the most important points is that 
$\mathcal{D}_2$ annihilates the mixed propagators 
$G^{+-}$ and $G^{-+}$, but it does not annihilate 
(anti-)time ordered one $G^{++(--)}$. 
Thus, we find that $\mathcal{D}_2G^{++}(\tau,\bm x; \tau_1,\bm x_1)=-\ri \delta(\tau-\tau_1) \delta^3(\bm x-\bm x_1)$, with the understanding that $\mathcal{D}_2$ acts on either $(\tau,\bm x)$ or $(\tau_1,\bm x_1)$. 
The Hermite conjugate of it yields that of the anti-time ordered products. Suppose we consider three-point 
correlation function $\langle\hat{\zeta}(\tau_*,\bm x_1)\hat{\zeta}(\tau_*,\bm x_2)\hat{\zeta}(\tau_*,\bm x_3)\rangle$ and treat them as $\zeta^+$. 
For concreteness, we formally write\footnote{One may consider the case where there are derivatives on $\zeta$, and even in such cases the following observations are applied.} $f(\zeta)=\lambda \zeta^2$. 
At first order in the interaction vertices, we would have contributions to the three-point function from 
the EOM term as
\begin{align}
 \langle\hat{\zeta}(\tau_*,\bm x_1)\hat{\zeta}(\tau_*,\bm x_2)\hat{\zeta}(\tau_*,\bm x_3)\rangle \supset  & -2\ri\lambda\int d\tau d^3\bm x\,\mathcal{D}_2G^{++}(\tau,\bm x;\tau_*,\bm x_1)G^{++}(\tau,\bm x;\tau_*,\bm x_2)G^{++}(\tau,\bm x;\tau_*,\bm x_3)\nonumber\\
&+2\ri\lambda\int d\tau d^3\bm x\,\mathcal{D}_2G^{-+}(\tau,\bm x;\tau_*,\bm x_1)G^{-+}(\tau,\bm x;\tau_*,\bm x_2)G^{-+}(\tau,\bm x;\tau_*,\bm x_3)+(\text{cyclic})\nonumber\\
=&-2\lambda\int d\tau d^3\bm x\,\delta(\tau-\tau_*)\delta(\bm x-\bm x_1)G^{++}(\tau,\bm x;\tau_*,\bm x_2)G^{++}(\tau,\bm x;\tau_*,\bm x_3)+(\text{cyclic})\nonumber\\
=&-2\lambda \langle \hat{\zeta}(\tau_*,\bm x_1)\hat{\zeta}(\tau_*,\bm x_2)\rangle\langle\hat{\zeta}(\tau_*,\bm x_1)\hat{\zeta}(\tau_*,\bm x_3)\rangle+(\text{cyclic}),
\end{align}
where ``(cyclic)'' means the terms with cyclic permutation of the indices $1,2,3$, and we have used the properties of the propagators mentioned above. Thus the EOM terms do not vanish, unlike the operator formalism. We also find the correspondence to the field redefinition method proposed by Maldacena~\cite{Maldacena:2002vr}. 
He proposed to introduce a new variable $\zeta=\zeta_n-f(\zeta_n)$, by which the quadratic part of the action cancels the EOM term\footnote{We note that the sign difference of $\zeta=\zeta_n-f(\zeta_n)$ in comparison 
to Ref.~\cite{Maldacena:2002vr} arises from the fact 
that we have defined $\delta S_{2}/\delta\zeta=\mathcal{D}_2\zeta$, which seems opposite to Ref.~\cite{Maldacena:2002vr}. }. 
However, from the observation here, we find that the EOM 
terms themselves 
play a role of the field redefinition at least at first order 
in the interaction vertices. 
It might be worth noting that the field redefinition in general leads to nontrivial Jacobian factor in the path integral, and therefore, such a technique might be cumbersome, whereas leaving the EOM terms may be tedious but straightforward\footnote{Nevertheless, we expect that any methods would be equally complicated starting from the second order in interaction vertices.}. 

We note the EOM terms always yield the time 
$\delta$-function that ``brings the interaction vertices to the same time plane on which operators are inserted''. 
Therefore, we have to be careful about the time-ordered product and the Heaviside step function $\theta(\tau)$ in the propagators \eqref{propagators}. We will take the following convention
\begin{align}
\theta(\tau)=\left\{\begin{array}{ll}1&\text{for }\tau>0\\ 1/2&\text{for }\tau=0\\ 0&\text{for }\tau<0\end{array} \right.\,.
\end{align}
This prescription keeps the reality of the correlation function of real fields\footnote{Independently of whether inserted operators are treated as $+$ or $-$ fields, the $\pm\ri$ from the vertices precisely cancel the $\mp\ri$ associated with $\mathcal{D}_2G^{++(--)}(x-x')=-(+)\ri \delta^4(x-x')$. Therefore, either choice yields the same result and the resultant contribution becomes real with the above prescription to the Heaviside $\theta$-function.}.

Let us turn to the discussion on the boundary terms. 
As we mentioned before, we set the final slice time $\tau_f$ to be later than $\tau_*$.
For concreteness, we assume $g(\zeta)=\lambda_2\zeta'\zeta^2$. 
Then, we find 
\begin{align}
    &\langle\hat{\zeta}(\tau_*,\bm x_1)\hat{\zeta}(\tau_*,\bm x_2)\hat{\zeta}(\tau_*,\bm x_3)\rangle\nonumber\\
&\supset  -2\ri\int d^3\bm x \biggl[\partial_{\tau} G^{++}(\tau,\bm x;\tau_*,\bm x_1)G^{++}(\tau,\bm x;\tau_*,\bm x_2)G^{++}(\tau,\bm x;\tau_*,\bm x_3)\nonumber\\
    &\qquad -\partial_{\tau} G^{-+}(\tau,\bm x;\tau_*,\bm x_1)G^{-+}(\tau,\bm x;\tau_*,\bm x_2)G^{-+}(\tau,\bm x;\tau_*,\bm x_3)+(\text{cyclic})\biggr]_{\tau\to\tau_f}\nonumber\\
   &=-2\ri \int dK\delta^3\left(\sum_{i=1}^3\bm k_i\right)e^{\ri(\sum_{i=1}^3\bm k_i\cdot\bm x_i)}\biggl[\partial_\tau G^{++}_{k_1}(\tau,\tau_*)G^{++}_{k_2}(\tau,\tau_*)G^{++}_{k_3}(\tau,\tau_*)\nonumber\\
   &\quad -\partial_{\tau} G^{-+}_{k_1}(\tau,\tau_*)G^{-+}_{k_2}(\tau,\tau_*)G^{-+}_{k_3}(\tau,\tau_*)+(\text{cyclic})\biggr]_{\tau\to\tau_f},\label{PIB}
\end{align}
where $\int dK=(2\pi)^{-6}\int d^3\bm k_1d^3\bm k_2d^3\bm k_3$. Since $G^{++}_{k_i}(\tau_f,\tau_*)=G^{-+}_{k_i}(\tau_f,\tau_*)=\zeta_{k_i}(\tau_f)\bar{\zeta}_{k_i}(\tau_*)$ for $\tau_f>\tau_*$, 
we find that the RHS of the above vanishes by cancellation between $++$ 
and $-+$, namely 
\begin{align}
\text{RHS of \eqref{PIB}}=0\,.
\end{align}
Here, we should emphasize that we have taken 
the inserted $\zeta$ operators as $+$-fields. Nevertheless, one can see that the cancellation 
occurs similarly even if we treat them to be 
$-$-fields\footnote{Essentially, in our prescription, the boundary interactions appear at either the most left or the most right for $\pm$-fields, and therefore, the $\pm$-vertices are not distinguished except for the overall sign difference, by which the contributions from $\pm$-Lagrangian cancel to each other.}. 
In this prescription, the boundary terms play no roles unlike the operator formalism where the EOM terms 
simply vanish.
This result is expected from the causality since the interaction on the future boundary must not affect the expectation values of operators on $\tau_*<\tau_f$.
As we will see in Sec.~\ref{examples}, the above prescription reproduces consistency relations of correlation functions of 
curvature perturbations.

Note however that there remains a puzzle: If we naively put the operators at the boundary time, i.e., $\tau_*=\tau_f$, the above cancellation seems not to occur and there would remain some nonvanishing contribution. Such a subtlety may arise because the boundary interaction appears at precisely the same time as the operators we evaluate. We will not discuss such a technical issue further as our prescription does not give such an ambiguity.

Let us comment on the boundary time in the operator formalism. In this formalism, the time evolution operator appears by translating the Heisenberg operator by the interaction picture operators. More specifically, if one wants to evaluate $\langle\hat{O}(\tau_*) \rangle$, where $\hat{O}(\tau_*)$ is supposed to be the Heisenberg picture operator, 
we rewrite it as
\begin{align}
\hat{O}(\tau_*)=\hat{U}^{\dagger}(\tau_*,\tau_0)\hat{O}^I(\tau_*)\hat{U}(\tau_*,\tau_0).
\end{align}
This implies that the ``boundary time'' is fixed by the operators that we evaluate. 
This feature differs from the path-integral case as we put operators by differentiating generating functionals with respect 
to the source functions, and therefore the ``boundary time'' is arbitrary. Such a difference may explain why the boundary terms are not taken to future infinity in the operator formalism\footnote{Of course, if we evaluate the operators at future infinity, this is not the case, and boundary operators also appear at infinity at least at the tree level.}. 

We also add a few comments on the relation with the 
wave functional formalism (see 
e.g.,\cite{Maldacena:2002vr,Harlow:2011ke,Pimentel:2013gza,Anninos:2014lwa}). In this formalism, there is also a notion of the boundary surface at the final time slice. Despite similarity to ours regarding the use of the path integral, in the wave functional formalism, the operators we evaluate are inserted on the final time surface, which differs from our criteria that we insert operators at the time well before the final time, which is why boundary surface does not effectively appear within our path integral formalism.

%%%%%%%%%%%%%%%%%%%%%%%%%%%%%%%%%%%%%%%%%%%%%%%%%%%%
\section{Example: three-point correlation functions 
and consistency relations in slow-roll inflation}
\label{examples}
%%%%%%%%%%%%%%%%%%%%%%%%%%%%%%%%%%%%%%%%%%%%%%%%%%%%%

In this section, we would like to check what the boundary 
and the EOM terms play the role in correlation functions. 
In particular, we focus on the consistency relations of curvature perturbations, which are important in proving the absence of the time-dependence of $\zeta$ that exits 
the Hubble horizon beyond a tree-level approximation~\cite{Pimentel:2012tw}. 

We assume a single-clock inflation model given by 
the action (\ref{action}) and take into account the Gibbons-Hawking-York action
$S_{\rm GHY}$ \cite{York:1972sj,Gibbons:1976ue}. 
The purpose of incorporating the term $S_{\rm GHY}$ is 
to make the variational principle well-defined in 
the bounded four-dimensional spacetime and to obtain 
the Lagrangian to be functions of $\zeta$ and $\zeta'$ 
(not including $\zeta''$).
In the unitary gauge, 
the cubic-order action of curvature perturbations 
consists of the following set of terms\footnote{In the following discussion, we focus only on the cubic terms. Strictly speaking, there are quadratic total time derivative terms, which seem to contribute as boundary terms. Nevertheless, such terms generally cancel 
when one converts the Lagrangian to the Hamiltonian. Therefore, one may safely neglect such terms.} 
(see e.g., Ref.~\cite{Collins:2011mz} for detailed derivation):
\begin{align}
S_{\rm bulk}=&M_{\rm Pl}^2\int d\tau d^3\bm x\,a^2\biggl[\epsilon^2\left(1-\frac12\epsilon\right)\zeta(\zeta')^2+\epsilon^2\zeta(\partial\zeta)^2+\frac12 \eta'\epsilon\zeta'\zeta^2-2\epsilon\zeta'\partial_k\zeta\partial_k(\partial^{-2}\zeta')+\frac12 \epsilon^3\zeta (\partial_k\partial_l\partial^{-2}\zeta')^2\biggr],\label{Sbulk}\\
S_{\partial}=&M_{\rm Pl}^2\int d^3\bm x\,
\biggl[-a^2\epsilon\eta\zeta'\zeta^2-\frac{a\epsilon}{H}\zeta(\zeta')^2-\epsilon^2\zeta'\zeta\partial^2\zeta+2a\epsilon^3\zeta^2\partial^2\zeta\nonumber\\
&\qquad+\frac{\epsilon}{2H^2}\zeta\partial_{k}\partial_l\zeta(\partial_k\partial_l\partial^{-2}\zeta')-\frac{1}{6aH}\zeta\{(\partial_k\partial_l\zeta)^2-(\partial^2\zeta)^2\}-\frac{a\epsilon^2}{2H}\zeta(\partial_k\partial_l\partial^{-2}\zeta')^2\biggr]\Biggr|_{\tau=\tau_f},\label{Sbd}\\
S_{\rm EOM}=&\int d \tau d^3\bm x\,\mathcal{D}_2\zeta 
\biggl[\frac14\eta\zeta^2+\frac{1}{aH}\zeta'\zeta-\frac{1}{4(aH)^2}\{(\partial\zeta)^2-\partial^{-2}\partial_k\partial_l(\partial_k\zeta\partial_l\zeta)\} \nonumber \\
&
+\frac{\epsilon}{aH}\{\partial_k\zeta\partial_k\partial^{-2}\zeta'-\partial^{-2}\partial_k\partial_l(\partial_k\zeta\partial_l\partial^{-2}\zeta)\}\biggr]\,,
\label{SEOM}
\end{align}
where $S_{\rm bulk}$, $S_{\partial}$, and $S_{\rm EOM}$ 
correspond to the contributions of the bulk, boundary, 
and EOM terms, respectively, and 
\be
\eta=\frac{\dot{\epsilon}}{H \epsilon}\,.
\ee
Note that the nonlocal operator $\partial^{-2}$ is understood in the sense of Fourier transform.

\subsection{Path-integral viewpoint}
\subsubsection{$\langle\zeta\zeta\zeta\rangle$}

Let us compute a three-point function $\langle\zeta(\tau_*,\bm x_1)\zeta(\tau_*,\bm x_2)\zeta(\tau_*,\bm x_3)\rangle$ in the path-integral 
formalism. As described in Sec.~\ref{pathintegral}, the boundary terms $S_\partial$ do not contribute to the tree-level correlation functions. Therefore, we are left with the bulk and the EOM terms. From the general consideration in Sec.~\ref{pathintegral}, at first order, the contraction with $\mathcal{D}_2\zeta$ and one of $\zeta(\tau_*,\bm x_i)$ ($i=1,2,3$) yields (anti-)time-ordered propagator that gives a $\delta$-function $\delta(\tau-\tau_*)\delta^3(\bm x-\bm x_i)$ by $\mathcal{D}_2$ acting on it, and the rest of $\zeta_{j\neq i}$ are contracted with remaining fields in the EOM terms, which reproduces the known result.
More specifically, by taking $\tau_*\to 0$ where $a\to\infty$, only the EOM term 
$S_{\rm EOM}\to\int d\tau d^3\bm x\, \mathcal{D}_2\zeta\left( \eta\zeta^2/4\right)$ survives\footnote{It is easy to confirm it even after taking contraction explicitly.}. 
Then, we find the contribution from the EOM 
term to be
\ba
\langle\zeta(\tau_*,\bm x_1)\zeta(\tau_*,\bm x_2)\zeta(\tau_*,\bm x_3)\rangle\biggr|_{\text{EOM}}
&\underset{\tau_*\to0} \sim&  
+\ri \int d\tau d^3\bm x\frac\eta2 \mathcal{D}_2G^{++}(\tau,\bm x;\tau_*,\bm x_1)G^{++}(\tau,\bm x;\tau_*,\bm x_2)(\tau,\bm x;\tau_*,\bm x_3)+({\rm cyclic})\nonumber\\
&=&
\frac{\eta(\tau_*)}{2}G^{++}(\tau_*,\bm x_1;\tau_*,\bm x_2)G^{++}(\tau_*,\bm x_1;\tau_*,\bm x_3)
+({\rm cyclic})\,,
\ea
where we have used $\mathcal{D}_2G^{++}(\tau,\bm x;\tau_*,\bm x_1)=-\ri\delta(\tau-\tau_*)\delta^3(\bm x-\bm x_1) $ and formally written the time-dependence of $\eta$ explicitly. We here emphasize again that there is no need for field redefinition, but we just straightforwardly compute the correlation function. 
The $-$ action yields just a vanishing contribution. 
Note that, if we assume $\zeta^-(\tau_*,\bm x_i)$ to be inserted at the boundary, the $-$ action instead contributes and the anti-time ordered propagator $G^{--}$ yields the same $\delta$ function with an opposite imaginary unit $+\ri \delta(\tau-\tau_*)\delta^3(\bm x-\bm x_i)$. 
This yields exactly the same result. 
The bulk contribution is computed in the same way as the known result studied in the literature.

To check the behavior of the consistency relation for the wavenumbers $k_3\ll k_1\sim k_2$, we show the final result expanded in terms of 
the slow roll parameters $\epsilon$ and $\eta$ 
(leaving $\tau_*<0$) in Fourier space, where we take the convention 
$\zeta(\tau,\bm k)\equiv \int d^3\bm x\, e^{\ri \bm k \cdot \bm x} \zeta(\tau,\bm x)$. 
Having fully computed the first-order contributions from the bulk and the EOM interactions\footnote{The boundary terms do not contribute as explained below \eqref{PIB} and hence need not be taken account of.} shown above to the three-point function and performed the Fourier transformation, we reach the result in the squeezed limit $k_3\ll k_1\sim k_2$ as
\begin{align}
\langle\zeta(\tau_*,\bm k_1)\zeta(\tau_*,\bm k_2)\zeta(\tau_*,\bm k_3)\rangle_c\biggr|_{\text{tot}}
\underset{k_3\ll k_1\sim k_2}{\longrightarrow}-\frac{H^4k_1^2\tau_*^2(2-\eta)}{16k_3^3k_1^3M_{\rm Pl}^6\epsilon^2}+\frac{H^4(\eta+2\epsilon)}{16k_3^3k_1^3M_{\rm Pl}^6\epsilon^2} +\mathcal{O}(k_3^{-2})\,,
\label{threepo}
\end{align}
where $\langle\ \cdot \ \rangle_c$ denotes the correlation function without overall $\delta$-function $(2\pi)^3\delta(\sum_i \bm k_i)$ and we have only shown the leading and next-to-leading order terms in slow-roll parameters. We note that the power spectrum under the slow-roll approximation is given by
\begin{align}
\langle{\zeta}(\tau_*, \bm k){\zeta}(\tau_*, -\bm k)\rangle_c\approx
\left\{\begin{array}{ll}
\dfrac{H^2}{8k^3M_{\rm Pl}^2\epsilon}(-k\tau_*)^{-\eta-2\epsilon}&~~\text{for }k\tau_*\to 0\,,\\
\dfrac{H^2k^2\tau_*^2}{4k^3M_{\rm Pl}^2\epsilon}& ~~\text{for }k\tau_* \gg 1\,,
\end{array}
\right.
\end{align}
where we have used the exact mode function 
\be
\zeta(\tau, k)=-\frac{\sqrt\pi}{2\sqrt2 \epsilon M_{\rm Pl}a}\sqrt{-\tau}H_{\nu}(-k\tau)\,,
\ee
with $\nu=3/2+\epsilon+\eta/2$.
Then, we find
\begin{align}
\langle\zeta(\tau_*,\bm k_1)\zeta(\tau_*,\bm k_2)\zeta(\tau_*,\bm k_3)\rangle_c\biggr|_{\text{tot}}\approx -\frac{d\log(k_1^3\langle{\zeta}(\tau_*, \bm k_1){\zeta}(\tau_*, -\bm k_1)\rangle_c)
}{d\log k_1}\langle{\zeta}(\tau_*, \bm k_3){\zeta}(\tau_*, -\bm k_3)\rangle_c\langle{\zeta}(\tau_*, \bm k_1){\zeta}(\tau_*, -\bm k_1)\rangle_c\,,\label{consistencyrelation}
\end{align}
which holds for $k_3\ll k_1\sim k_2$ and for either $k_1\tau_*\gg1$ or $k_1\tau_*\ll1$. 
This is the well-known Maldacena's consistency relation \cite{Maldacena:2002vr}. 
It would be noteworthy that the leading-order contribution
to the above for the sub-Hubble mode $k_1\tau_*\gg 1$ 
comes from the EOM term rather than the bulk terms. 

\subsubsection{$\langle\dot\zeta\dot\zeta\zeta\rangle$}

Let us consider another nontrivial example of the three-point correlation function. It would be illustrative to consider the consistency relation of $\langle \dot{\hat{\zeta}}(\tau_*,\bm k_1)\dot{\hat{\zeta}}(\tau_*,\bm k_2)\hat{\zeta}(\tau_*,\bm k_3)\rangle$. Since we expect that the correlation function simply vanishes when $\tau_*\to 0$ or all the modes are in 
the super-Hubble regime, we only focus on the case $k_3\ll k_1\sim k_2$ while $k_1\tau_*=\text{finite}$. The computation of the bulk contribution is rather straightforward since nothing really changes from the standard calculation except that the inserted operators contain extra time derivatives. Hence we focus only on the contribution from EOM terms. If one notices that the 
tree-level contribution of the EOM terms can be reproduced by the well-known field redefinition procedure, the computation is also very straightforward. Nevertheless, with the application to the higher-order corrections including loops in mind, we give some details of the computation leaving the EOM terms as they are. Firstly, we start with the position-space correlation function $\langle\dot{\hat{\zeta}}(\tau_*,\bm x_1)\dot{\hat{\zeta}}(\tau_*,\bm x_2)\hat{\zeta}(\tau_*,\bm x_3)\rangle$ instead. 
The contractions yield three different contributions $F_1, F_2, F_3$, 
which are given, respectively, by\footnote{$F_i$'s correspond 
to the contributions from the contraction of $\mathcal{D}_2\zeta$ in the EOM interaction and the external field $\zeta(\bm x_i)$.}
\begin{align}
F_1=&\frac{\ri}{a^2(\tau_*)}\int d\tau d^3\bm x\, 
(-\ri)\partial_{\tau_2}\delta(\tau-\tau_2)\delta^3(\bm x-\bm x_1)\biggl[\frac\eta2\partial_{\tau_1}G^{++}(\tau_1,\bm x_2;\tau,\bm x)G^{++}(\tau_1,\bm x_3;\tau,\bm x)\nonumber\\
&+\frac{1}{aH}\partial_\tau\{\partial_{\tau_1}G^{++}(\tau_1,\bm x_2;\tau,\bm x)G^{++}(\tau',\bm x_3;\tau,\bm x)\}-\frac{1}{2(aH)^2}\partial_{\tau_1}\partial_kG^{++}(\tau_1,\bm x_2;\tau,\bm x)\partial_kG^{++}(\tau_1,\bm x_3;\tau,\bm x)\nonumber\\
&+ \frac{1}{2(aH)^2}\partial^{-2}\partial_k\partial_l\{\partial_{\tau_1}\partial_kG^{++}(\tau_1,\bm x_2;\tau,\bm x)\partial_lG^{++}(\tau_1,\bm x_3;\tau,\bm x)\} \nonumber\\
&
-\frac{\epsilon}{aH}\partial^{-2}\partial_k\partial_l\partial_\tau\{\partial_{\tau_1}\partial_kG^{++}(\tau_1,\bm x_2;\tau,\bm x)\partial_lG^{++}(\tau_1,\bm x_3;\tau,\bm x)\}\nonumber\\
&+\frac{\epsilon}{aH}\{\partial_{\tau_1}\partial_k\partial^{-2}G^{++}(\tau_1,\bm x_2;\tau,\bm x)\partial_kG^{++}(\tau_1,\bm x_3;\tau,\bm x)+\partial_{\tau_1}\partial_kG^{++}(\tau_1,\bm x_2;\tau,\bm x)\partial_k\partial^{-2}G^{++}(\tau_1,\bm x_3;\tau,\bm x)\}
 \biggr]_{\tau_{1,2}\to\tau_*},\\
&  F_2=F_1|_{\bm x_1\leftrightarrow \bm x_2},
\end{align}
and 
\begin{align}
F_3=&\frac{\ri}{a^2(\tau_*)}\int d\tau d^3\bm x (-\ri)\delta(\tau-\tau_1)\delta^3(\bm x-\bm x_3)\biggl[\frac\eta2\partial_{\tau_1}G^{++}(\tau_1,\bm x_1;\tau,\bm x)\partial_{\tau_1}G^{++}(\tau_1,\bm x_2;\tau,\bm x)\nonumber\\
&+\frac{1}{aH}\partial_\tau\{\partial_{\tau_1}G^{++}(\tau_1,\bm x_1;\tau,\bm x)\partial_{\tau_1}G^{++}(\tau_1,\bm x_2;\tau,\bm x)\}-\frac{1}{2(aH)^2}\partial_{\tau_1}\partial_kG^{++}(\tau_1,\bm x_1;\tau,\bm x)\partial_{\tau_1}\partial_kG^{++}(\tau_1,\bm x_2;\tau,\bm x)\nonumber\\
&+ \frac{1}{2(aH)^2}\partial^{-2}\partial_k\partial_l\{\partial_{\tau_1}\partial_kG^{++}(\tau_1,\bm x_1;\tau,\bm x)\partial_{\tau_1}\partial_lG^{++}(\tau_1,\bm x_2;\tau,\bm x)\}\nonumber\\
&+\frac{\epsilon}{aH}\{\partial_{\tau_1}\partial_k\partial^{-2}G^{++}(\tau_1,\bm x_1;\tau,\bm x)\partial_{\tau_1}\partial_kG^{++}(\tau_1,\bm x_2;\tau,\bm x)+\partial_{\tau_1}\partial_kG^{++}(\tau_1,\bm x_1;\tau,\bm x)\partial_{\tau_1}\partial_k\partial^{-2}G^{++}(\tau_1,\bm x_2;\tau,\bm x)\}\nonumber\\ 
&-\frac{\epsilon}{aH}\partial^{-2}\partial_k\partial_l\partial_\tau\{\partial_{\tau_1}\partial_kG^{++}(\tau_1,\bm x_1;\tau,\bm x)\partial_{\tau_1}\partial_lG^{++}(\tau_1,\bm x_2;\tau,\bm x)\} \biggr]_{\tau_1\to\tau_*}.
\end{align}
We have shown (possibly unnecessary) whole terms just for completeness. It is straightforward to perform the integrals in each quantity, but only the $\tau$ integration in $F_{1,2}$ needs a little care due to the derivative acting on the $\delta$-function. 
Since the $\tau_2$ derivative is acting 
on $\delta$-function, we obtain
\begin{align}
F_1=+\partial_\tau\biggl[\frac\eta2\partial_{\tau_1}G^{++}(\tau_1,\bm x_2;\tau,\bm x)G^{++}(\tau_1,\bm x_3;\tau,\bm x)+\cdots\biggr]\,.
\end{align}
One may check that this is consistent with the result of field redefinition. Thus, even if inserted operators contain derivatives, one can straightforwardly compute correlation functions without the field redefinition procedure. 

After straightforward but messy calculations, it turns out that the bulk contribution only gives the subleading contribution in the squeezing limit $k_3\ll k_1\sim k_2$, and a specific EOM term in the Lagrangian, ${\cal{L}}_{\rm EOM} \supset ({\cal{D}}_2\zeta) \zeta'\zeta/aH$ yields the leading contribution 
\ba
& &
\langle \dot{\hat{\zeta}}(\tau_*,\bm k_1)\dot{\hat{\zeta}}(\tau_*,\bm k_2)\hat{\zeta}(\tau_*,\bm k_3)\rangle_c \underset{k_3\ll k_1\sim k_2}{\approx}
F_1+F_2\approx -\frac{H^6k_1\tau_*^4}{4k_3^3M_{\rm Pl}^4\epsilon^2} \nonumber \\
\hspace{-6cm}
~~~~~& &\approx 
-\frac{d\log(k_1^3\langle{\dot{\zeta}}(\tau_*, \bm k_1){\dot{\zeta}}(\tau_*, -\bm k_1)\rangle_c)
}{d\log k_1}\langle{\zeta}(\tau_*, \bm k_3){\zeta}(\tau_*, -\bm k_3)\rangle_c\langle{\dot{\zeta}}(\tau_*, \bm k_1){\dot{\zeta}}(\tau_*, -\bm k_1)\rangle_c\,,
\ea
which again corresponds to the consistency relation. Thus, we have shown that the EOM terms are responsible for the consistency relation in the squeezed limit.

We nevertheless emphasize that whether the bulk or the EOM terms are responsible for the consistency relation 
does not really make sense. We can change 
the bulk, boundary, and EOM terms according to our preference by using the identity shown in Appendix~\ref{ambiguityId}.

It would be worth noting that, in the operator formalism, the Heisenberg operator $\dot{\hat{\zeta}}$ differs from that in the interaction picture $\dot{\hat{\zeta}}^I$ as 
discussed, e.g., in Ref.~\cite{Pimentel:2012tw}, and accordingly, one has to take it into account in evaluating correlation functions. This type of difference does not appear in the path-integral case since the original source function is formally of the Heisenberg operator, and therefore, the correction appearing in the operator formalism does not appear. In other words, such a correction would be automatically taken into account perturbatively. This is the reason why naively evaluated 
$\langle\dot\zeta\dot\zeta\zeta\rangle$ shows the correct soft behavior without any corrections associated with the difference between the Heisenberg and the interaction pictures.

\subsection{Operator formalism viewpoint}

We briefly discuss the consistency relation within the operator formalism. As we have discussed, the boundary terms contribute 
to the computation of the three-point correlation function, 
whereas the EOM terms trivially vanish in the operator 
formalism of the interaction picture. Since the role of the boundary operators was already discussed in \cite{Arroja:2011yj,Burrage:2011hd} for soft modes, we only focus on the consistency relation of
$\langle \zeta\zeta\zeta\rangle$ 
with the first two fields in the sub-Hubble regime 
($k_1 \tau_* \gg 1$ and $k_2 \tau_* \gg 1$), while the 
other is in the super-Hubble regime ($k_3 \tau_* \ll 1$). 
In the corresponding squeezed limit ($k_3\ll k_1\sim k_2$), 
the leading-order term can be found from the contribution of a single boundary operator\footnote{This is not 
a Hamiltonian operator as it is dimensionless.}
\be
\hat{H}_\partial=-\int d^3\bm x \hat{\mathcal{L}}_\partial
\supset \int d^3\bm x \frac{M_{\rm Pl}^2a\epsilon}{H}\hat{\zeta}'\hat{\zeta}\hat{\zeta}'\biggl|_{\tau=\tau_*}\,,
\ee
where we have symmetrized the interaction such that the operator becomes Hermitian. The boundary operator is located at the time where the operators are inserted, which enables us to use the equal-time canonical commutation relations. 
For $\langle \zeta\zeta\zeta\rangle$, we find
\ba
\langle\hat\zeta(\bm x_1)\hat\zeta(\bm x_2)\hat\zeta(\bm x_3)\rangle_c &\supset&
\frac{\ri M_{\rm Pl}^2a\epsilon}{H}\int d^3\bm x [\hat{\zeta}_{\bm x}'\hat{\zeta}_{\bm x}\hat{\zeta}_{\bm x}',\hat{\zeta}_1\hat{\zeta}_2\hat{\zeta}_3] \nonumber\\
&=&\frac{1}{2aH}\left[\langle(\hat{\zeta}'_1\hat\zeta_1+\hat{\zeta}_1\hat\zeta'_1)\hat{\zeta}_2\hat{\zeta}_3\rangle+\hat{\zeta}_1(\hat\zeta'_2\hat\zeta_2+\hat\zeta_2\hat\zeta'_2)\hat\zeta_3+\hat\zeta_1\hat\zeta_2(\hat\zeta'_3\hat\zeta_3+\hat\zeta_3\hat\zeta'_3)\right],\label{zeta3bd}
\ea
where we have used abbreviation $\hat{\zeta}(\bm x)=\hat\zeta_{\bm x}$, and $\hat{\zeta}_i=\hat\zeta(\bm x_i)$ and omitted the time $\tau=\tau_*$. Evaluating the Fourier transform 
of Eq.~\eqref{zeta3bd} yields
\begin{align}
\langle\hat\zeta(\bm k_1)\hat\zeta(\bm k_2)\hat\zeta(\bm k_3)\rangle_c\supset& \frac{1}{2aH}\frac{d}{d\tau}\left[|\zeta_{k_2}(\tau)|^2|\zeta_{k_3}(\tau)|^2+(\text{cyclic})\right]_{\tau\to\tau_*}\underset{k_3\ll k_1\sim k_2}{\longrightarrow}-\frac{H^4 k_1^2\tau_*^2}{8M_{\rm Pl}^4\epsilon^2 k_1^3k_3^3} +\mathcal{O}(k_3^{-1})\,,
\end{align}
which coincides with the leading-order slow-roll 
contribution to Eq.~(\ref{threepo}) for $k_1\tau_*\sim k_2\tau_*\gg1$.
Thus, as the EOM terms are required in the path-integral formalism, the boundary operators are necessary to reproduce physically important properties such as the 
consistency relations.

In the above calculation, the two modes are 
in the sub-Hubble regime, with the other mode in the 
super-Hubble regime. Performing a similar calculation where all three modes are in the super-Hubble regime, it follows that 
both the boundary and the bulk terms are necessary to 
prove the consistency relation. 

%%%%%%%%%%%%%%%%%%%%%%%%%%%%%%%%%%%%%%%%%%%%%%%%%
\section{Comment on computation of higher-order terms}\label{Higherorder}
%%%%%%%%%%%%%%%%%%%%%%%%%%%%%%%%%%%%%%%%%%%%%%%%%

In this section, we give a few comments on 
the computation of higher-order corrections such as loops. 
As is clear from our discussion so far, the boundary (EOM) terms are not negligible for the operator (path-integral) formalism. In particular, we are mostly concerned with the large-scale modes generated and exiting the Hubble radius during inflation, which is expected to be ``decoupled''. As we have seen, for the consistency relation to be satisfied, we need to correctly include the boundary and the EOM terms, even for 
higher-order corrections. 

In operator formalism, it is rather straightforward to take the boundary terms into account by simply leaving 
it in the interaction Hamiltonian. However, we should notice that it does not look like boundary terms at higher order. 
More specifically, at second order, the expectation value 
of an operator $\hat{\mathcal{O}}(\tau_*)$ receives 
\ba
\langle \hat{\mathcal{O}}(\tau_*)\rangle 
&\supset& \ri^2\int^{\tau_*}_{-\infty} d\tau_1\int^{\tau_1}_{-\infty} d\tau_2\left\langle \left[\hat{H}^I_{\rm bulk}(\tau_2)+\frac{d}{d\tau_2}\hat{g}^I(\tau_2),\left[\hat{H}^I_{\rm bulk}(\tau_1)+\frac{d}{d\tau_1}\hat{g}^I(\tau_1),\hat{\mathcal O}^I(\tau_*)\right]\right]\right\rangle \nonumber\\
&=&-\int^{\tau_*}_{-\infty} d\tau_1\int^{\tau_1}_{-\infty} d\tau_2\left\langle \left[\hat{H}^I_{\rm bulk}(\tau_2),\left[\hat{H}^I_{\rm bulk}(\tau_1)+\frac{d}{d\tau_1}\hat{g}^I(\tau_1),\hat{\mathcal O}^I(\tau_*)\right]\right]\right\rangle 
\nonumber\\
&& -\int^{\tau_*}_{-\infty} d\tau_1\left\langle \left[\hat{g}^I(\tau_1),\left[\hat{H}^I_{\rm bulk}(\tau_1)+\frac{d}{d\tau_1}\hat{g}^I(\tau_1),\hat{\mathcal O}^I(\tau_*)\right]\right]\right\rangle\,,
\label{BDsecond}
\ea
where we have assumed $\hat{g}^I(\tau)\to 0$ as 
$\tau\to-\infty$ for the equality in the second line. 
The point is that the total time-derivative term inside the commutator cannot be simply performed. 
In the second line on the RHS of Eq.~\eqref{BDsecond}, the boundary terms ``behave'' like a bulk term in general, which would be understood, e.g., from the fact that the operator 
\begin{align}
\hat{u}(\tau_1)\frac{d}{d\tau_1}\hat{g}^I(\tau_1)
\end{align}
is no longer a total time derivative of some operator in general, where $\hat{u}(\tau_1)\equiv\int^{\tau_1}_{-\infty} d\tau_2\hat{H}^I_{\rm bulk}(\tau_2)$. Therefore, it would be practical to think of the total time derivative operator to be a part of the bulk Hamiltonian. As long as we keep the total time-derivative terms, we are free to throw away the EOM terms within the interaction picture operator formalism for the reason explained before. In this sense, despite complexities, it is rather straightforward to compute higher-order corrections.

It might be useful to also note that the last term 
appearing in Eq.~\eqref{BDsecond} can be rewritten as
\begin{align}
-\int^{\tau_*}_{-\infty} d\tau_1\left\langle \left[\hat{g}^I(\tau_1),\left[\hat{H}^I_{\rm int}(\tau_1),\hat{\mathcal O}^I(\tau_*)\right]\right]\right\rangle=
-\int^{\tau_*}_{-\infty} d\tau_1\left\langle\left[\hat{H}^I_{4}(\tau_1),\hat{\mathcal O}^I(\tau_*)\right]- \left[\hat{H}^I_{\rm int}(\tau_1),\left[\hat{g}^I(\tau_1),\hat{\mathcal O}^I(\tau_*)\right]\right]\right\rangle,
\end{align}
where we have defined an effective quartic interaction
\begin{align}
\hat{H}_4^I(\tau_1)=\left[\hat{g}^I(\tau_1),
\hat{H}_{{\rm int}}^I(\tau_1)\right]\,.
\end{align}
Since this operator is made of an equal-time commutator, 
only the part containing the canonical commutation relation gives spatial $\delta$-function, with which the interactions become quartic contact terms. Here, we should be careful about the corrections associated with the interactions with time-derivatives, which makes the relation between $\zeta'$ and canonical momentum $\pi_\zeta$ nonlinear. In the above discussion, we have not taken into account such corrections. 
In Ref.~\cite{Braglia2024}, it was shown that such corrections cancel some part of the effective quartic terms mentioned above, which implies that one must carefully define the effective Hamiltonian in the presence 
of total time derivative terms (or more generally, in the system 
with time-derivative interactions).\footnote{We would like to thank the authors of \cite{Braglia2024} for explaining this point to us.}

In the path-integral formalism, we have to deal with the EOM terms rather than the boundary terms. As we have shown, their role is equivalent to the field redefinition, but an advantage of keeping it as interaction terms rather than field redefinition is that we may straightforwardly compute correlation functions without careful treatment of the path-integral measure, which is required when one performs field redefinition, in general. More specifically, we often use the field redefinition of the form $\zeta \to \zeta+f(\zeta)$, which leads to the functional Jacobian $\left|{\rm Det}\left(\delta^4(x-y)+\delta f(\zeta(x))/\delta \zeta(y) \right)\right|$. Of course, such an advantage would be a small technical issue, and in the case of the S-matrix in flat spacetime, it is known that field redefinition does not change the result~\cite{Chisholm:1961tha,Kamefuchi:1961sb} and the nontrivial Jacobian factor vanishes by working with dimensional regularization scheme~\cite{Arzt:1993gz} (except for chiral anomalies). 
It is not immediately clear whether a similar argument applies to the case of the cosmological correlators as the terms removed by field redefinition contain $\zeta'$ as well as nonlocal operators containing $\partial^{-2}$.  As far as we know, such an issue has not been addressed in the literature and it is worth investigating further. We will leave it for future work. At least, we argue that such a subtle issue does not appear as long as we leave EOM terms since we do not need any field redefinition.

We give a few comments on higher-order corrections including the EOM terms. As already discussed, the EOM terms contribute to correlation functions only when the field with the EOM derivative operator is contracted 
by (anti-)time-ordered propagators, which yields the spacetime $\delta$-function. 
For instance, we consider the contraction of the following 
two interaction vertices,
\begin{align}
& +\ri \int d\tau_1d^3\bm x_1\, 
f\left(\zeta^+(\tau_1,\bm x_1)\right)\contraction{\mathcal{D}_2}{\zeta^+}{(\tau_1,\bm x_1)\times(+\ri) \int d\tau_2d^3\bm x_2 \sum}{g}\mathcal{D}_2\zeta^+(\tau_1,\bm x_1)\times(+\ri) \int d\tau_2d^3\bm x_2 \sum g\left(\zeta^+(\tau_2,\bm x_2)\right) \nonumber\\
&\supset -\prod_{i=1}^2\int d\tau_id^3\bm x_i\, 
\mathcal{D}_2G^{++}(\tau_1,\bm x_1;\tau_2,\bm x_2)f(\zeta^+(\tau_1,\bm x_1))\frac{\delta g}{\delta\zeta^+}(\tau_2,\bm x_2)=+\ri\int d\tau_1d\bm x_1 f(\zeta^+)\frac{\delta g}{\delta\zeta^+}\,.
\end{align}
If they are originally cubic interactions, the above term is a contact four-point interaction like the case of the boundary 
and the bulk contraction in the operator formalism mentioned above. Note that, in the above expression, we have taken just a single contraction, which is just a part of the full contractions. Thus, the inclusion of the EOM term to the second order effectively yields ``tree level'' contributions. This is simply expected from the viewpoint of the field redefinition as well. The appearance of the quartic contact terms is observed in the operator formalism as discussed above, and it would be related to the quartic terms discussed in Ref.~\cite{Pimentel:2012tw}, which play a crucial role 
in reproducing the consistency relation and proving the 
time-independence of curvature perturbations.

As we have shown explicitly, with the set of interaction terms~\eqref{Sbulk}-\eqref{SEOM}, the boundary or the EOM terms are necessary to reproduce the consistency relations correctly, and therefore, one has to include them properly, otherwise one may find incorrect ultraviolet behavior of the loop corrections, which would be relevant to the claim 
made in Refs.~\cite{Fumagalli:2023hpa,Tada:2023rgp}.

%%%%%%%%%%%%%%%%%%%%%%%%%%%%%%%%%%%
\section{Summary}\label{summary}
%%%%%%%%%%%%%%%%%%%%%%%%%%%%%%%%%%%%

In this paper, we have discussed how to deal with the boundary 
and EOM terms appearing in cosmological QFT calculations. 
In operator formalism, the EOM term trivially vanishes, but the boundary terms contribute to the correlation function 
in a nontrivial manner. On the other hand, in the path-integral formalism, both of them seem to contribute to the correlation function, but (at least) at the lowest order, the boundary terms do not contribute under our prescription where the final time slice $\tau_f$ is set later than $\tau_*$, the time of the operators whose expectation is evaluated. 

The crucial difference between the two formalisms boils down to how derivatives in the interactions are treated. In the operator formalism, the derivatives directly act on operators and we evaluated the differentiated operators in the correlation functions, whereas in the path-integral formalism, we evaluate all by the free propagators and the derivatives act on the propagators rather than operators. Either method would give a correct result, so long as we carefully treat the boundary and the EOM terms. These observations resolve some misleading statements, particularly for the EOM terms that 
they trivially vanish. 
With the understanding developed in this work, it would be rather straightforward to consider higher-order corrections, 
for which we have also given a few comments. 
In particular, it has been known that consistency relations play crucial roles in proving the conservation of super-Hubble 
curvature perturbations against possible time-dependence from loop corrections (see e.g.,~\cite{Pimentel:2012tw}), and as we have partially seen, the boundary or the EOM terms may become responsible for the consistency relations of sub-Hubble modes, 
which would contribute to the ultraviolet part of the loop calculations. 
As long as the boundary and the EOM terms are correctly treated, we would be able to compute arbitrarily higher-order corrections precisely. 

%\cy{In our separate work~\cite{Kawaguchi:2024rsv}, we have indeed proved the absence of large one-loop corrections to the power spectrum in transient ultra-slow-roll inflation models.
%In the proof, the consistency relations play significant roles since the decoupling of super-horizon $\zeta$ is a property of the spacetime symmetry.
%As shown in this paper, in order for the consistency relations to hold, it is necessary to include the boundary terms for the operator formalism or the EOM terms for the path integral formalism.
%Therefore, the boundary and the EOM terms we have addressed in this work should never be missed in any inflationary models.
%}

Even in transient ultra-slow-roll models, the consistency relations should hold for super-Hubble modes that exit the Hubble horizon well before the ultra-slow-roll stage. If it is the case, we expect the decoupling of such $\zeta$ since it is a consequence of residual spacetime symmetry. In this sense, proving consistency relations as we perform in this work is a key to understand one-loop corrections to super-Hubble curvature perturbations. In our separate work~\cite{Kawaguchi:2024rsv}, 
we have indeed shown that EOM terms are necessary to prove the consistency relations in transient ultra-slow-roll inflation models, which accordingly proves the absence of large one-loop corrections to the large-scale 
power spectrum. Therefore, the boundary and EOM terms we have addressed in this work should never be missed in any inflationary models. 

It was also shown that the same consistency relation as 
the Maldacena's one 
holds for more general slow-variation single-field inflation 
in Horndeski's theories with second-order field equations of 
motion \cite{DeFelice:2013ar} (see also Ref.~\cite{DeFelice:2011uc}), by picking up dominant boundary 
and EOM terms in the cubic action with the interaction picture.  
It will be of interest to formally apply our prescription
of the operator and path-integral formalisms to 
such more general theories to provide a more rigorous
proof for the consistency relations.

%%%%%%%%%%%%%%%%%%%%%%%%%%%%%%%%%%%%%%%%%%%%%%%%%%%%%%%%%%%%%%%%%
\section*{Note added}
While completing this work, we became aware of 
a related work~\cite{Braglia2024}. 
In this paper, the authors discussed the same issue regarding 
the total time derivative and EOM terms in great detail, 
which includes the points we have not addressed in this work. 
We mostly agree with their results, but our proposal regarding the boundary terms within the path-integral formalism seems rather simple as our argument shows that the boundary terms can be neglected within it by causality. Therefore, it is not so cumbersome to use the known 
path-integral formalism as one only needs to take the EOM terms. 
In Ref.~\cite{Braglia2024}, a new canonical transformation approach is proposed. In our opinion, we are not yet sure if the canonical transformation leads to quantum anomalies. Conservatively, it is reasonable to leave EOM or boundary terms as they are to proceed 
higher-order computations in perturbation theory. 
%%%%%%%%%%%%%%%%%%%%%%%%%%%%%%%%%%%%%%%%%%%%%%%%%%%%%%%%%%%

%%%%%%%%%%%%%%%%%%%%%%%%%%%%%%%%%%%%%%%%%%%%%%%%%%%%%%%%%%%
\section*{Acknowledgements}
We thank Tomohiro Fujita, Mohammad Ali Gorji, Jason Kristiano, 
and Yuichiro Tada for useful discussions. 
We would also like to thank Matteo Braglia and Lucas Pinol 
for useful comments and discussions. 
ST is supported by the Grant-in-Aid for Scientific 
Research Fund of the JSPS No.~22K03642 
and Waseda University Special Research Project 
No.~2023C-473. YY is supported by Waseda University 
Grant for Special Research Projects (Project number: 2023C-584).
%%%%%%%%%%%%%%%%%%%%%%%%%%%%%%%%%%%%%%%%%%%%%%%%%%%%%%%%%%%

\appendix

%%%%%%%%%%%%%%%%%%%%%%%%%%%%%%%%%%%%%%%%%%%%%%%%%%
\section{Ambiguity of the boundary and the EOM terms}\label{ambiguityId}
%%%%%%%%%%%%%%%%%%%%%%%%%%%%%%%%%%%%%%%%%%%%%%%%%%

The boundary and the EOM terms can be arbitrarily changed by the following procedure: One finds the following identity
\begin{align}
0= \int d^4x\left[ {\mathcal D}_2\zeta f(\zeta)-\frac{d}{dt}\left(2a^3\epsilon\dot\zeta f(\zeta)\right)+2a^3\epsilon\dot\zeta \dot{f}(\zeta)-2a\epsilon\,\partial\zeta\partial f(\zeta)\right]\,,
\end{align}
where we have neglected the total spatial derivative terms. Adding such a ``zero'' to the action changes the bulk, boundary, and EOM terms. Such ambiguity can be used to change the interaction terms. As we have shown in the main text, any of them cannot be neglected in general. 

%%%%%%%%%%%%%%%%%%%%%%%%%%%%%%%%%%%%%%%%%%%%%%%%%%%%%
\section{Remarks on boundary terms}\label{noteonBT}
%%%%%%%%%%%%%%%%%%%%%%%%%%%%%%%%%%%%%%%%%%%%%%%%%%%%%

We discuss some subtleties of the total time-derivative term 
in the Hamiltonian. As an illustration, we use the following simple time-dependent Hamiltonian in quantum mechanics:
\begin{align}
\hat{H}[\hat{\phi}(t),\hat{\pi}(t);t]=\hat{H}_0[\hat{\phi}(t),\hat{\pi}(t);t]+\hat{A}[\hat{\phi}(t),\hat{\pi}(t);t]+\frac{d}{dt}\hat{B}[\hat{\phi}(t),\hat{\pi}(t);t]\,,
\end{align}
where $\hat\pi$ is the canonical momentum operator of $\hat{\phi}$,  $\hat{H}_0[\hat{\phi}(t),\hat{\pi}(t);t]$ is the free part of the Hamiltonian, and $\hat{A}[\hat{\phi}(t),\hat{\pi}(t);t]$ is the (bulk) interaction term. 
We have assumed that the Hamiltonian has 
explicit time-dependence, e.g., in the couplings as the cosmological models. 
The above definition of the Hamiltonian already has a subtlety: The Hamiltonian 
contains the total time derivative of $\hat{B}$, but its definition depends on the Hamiltonian 
itself since
\begin{align}
\frac{d}{dt}\hat{B}[\hat{\phi}(t),\hat{\pi}(t);t]=\hat{U}^{-1}_S(t,t_0)\frac{\partial}{\partial t}\hat{B}[\hat{\phi}(t_0),\hat{\pi}(t_0);t]\hat{U}_S(t,t_0)+\ri \left[\hat{H}[\hat{\phi}(t),\hat{\pi}(t);t], \hat{B}[\hat{\phi}(t),\hat{\pi}(t);t]\right]\,,
\end{align}
where $\hat{U}_S(t,t_0)$ is the Schr\"odinger picture time-evolution operator satisfying
\begin{align}
\ri\frac{d}{dt} \hat{U}_S(t,t_0)=
\hat{H}[\hat{\phi}(t_0),\hat{\pi}(t_0);t]\hat{U}_S(t,t_0)\,,
\end{align}
where $t_0$ denotes the initial time. In defining the Schr\"odinger picture Hamiltonian 
$\hat{H}[\hat{\phi}(t_0),\hat{\pi}(t_0);t]\equiv \hat{h}(t)$, the consistency with the Heisenberg 
picture requires that 
\begin{align}
\hat{h}(t)=\hat{h}_0(t)+\hat{A}_0(t)+\frac{\partial}{\partial t}\hat{B}_0(t)+\ri[\hat{h}(t),\hat{B}_0(t)]\,,
\end{align}
where $\hat{h}_0(t)=\hat{H}_0[\hat{\phi}(t_0),\hat{\pi}(t_0);t]$, $\hat{A}_0(t)=\hat{A}[\hat{\phi}(t_0),\hat{\pi}(t_0);t]$, and $\hat{B}_0(t)=\hat{B}[\hat{\phi}(t_0),\hat{\pi}(t_0);t]$. 
We note that the last term is necessary to reproduce the total time derivative term in the original Hamiltonian in the Heisenberg picture $\hat{H}=\hat{U}_S^{-1}\hat{h}\hat{U}_S$. One may use the interaction picture based on the free Hamiltonian $\hat{H}_0$, but one would immediately find that the interaction Hamiltonian is yet dependent on the full interaction Hamiltonian. 
Thus, the time evolution of either states or operators must be in a self-consistent way. 
As far as we know, there is no way to define a Hamiltonian without a term containing 
the Hamiltonian itself. 

The above issue appears not only in the Hamiltonian with a boundary term but the system with terms containing 
time-derivatives on the canonical variables $\phi(t),\pi(t)$. This is somewhat reasonable 
as higher time derivatives lead to ghost poles in general, which gives rise to additional degrees of freedom in the theory. We expect that such higher derivative terms do not appear in ultraviolet complete theories but only in their effective field theories. 
In particular, there are no terms containing higher-derivative terms in general relativity
with minimal matter couplings, and there is a Hamiltonian that has no subtleties discussed above, and one has to define 
a theory with it. 

Suppose that our Hamiltonian does not have the total time derivative terms posing the above issue. 
In this case, one can first quantize the system and move to the 
interaction picture, and then we may integrate the interaction Hamiltonian by part, 
which effectively induces the total time derivative terms as is the case of metric perturbations 
in general relativity. Nevertheless, it should be noted that, 
even if higher-derivative terms are absent, it is nontrivial to find a full Lagrangian 
if the Lagrangian contains powers of $\dot\phi$ since its canonical conjugate 
is no longer $\pi_\phi \propto \dot\phi$. In such a case, one has to carefully 
define the Hamiltonian.

%%%%%%%%%%%%%%%%%%%%%%%%%%%%%%%%%%%%%%%%
\bibliographystyle{mybibstyle}
\bibliography{bib}

\end{document}